\documentclass[aps,pra,preprint,groupedaddress,showpacs]{revtex4}
\usepackage[dvips]{graphicx}

\begin{document}

\title{Effect of collision dephasing on atomic evolutions in a high-\textit{Q%
} cavity}
\author{T.W.~Chen and P.T.~Leung}
\affiliation{Department of Physics, The Chinese University of Hong Kong, \\
Shatin, Hong Kong SAR, China}
\date{\today }

\begin{abstract}
The decoherence mechanism of a single atom inside a high-\textit{Q} cavity
is studied, and the results are compared with experimental observations
performed by M. Brune \textit{et al}. [Phys. Rev. Lett. \textbf{76}, 1800
(1996)]. Collision dephasing and cavity leakage are considered as the major
sources giving rise to decoherence effect. In particular, we show that the
experimental data can be fitted very well by assuming suitable values of
collision Stark shifts and dark count rate in the detector.
\end{abstract}

\maketitle


A two-level atom interacting with a single cavity mode, the Jaynes-Cummings
model (JCM) \cite{JC}, is possibly the most well studied system in quantum
optics and has the potentiality to constitute the basic building block of
quantum computers \cite{computer}. Among all the intriguing phenomena
related to the JCM, the oscillation in the atomic inversion probability ---
the Rabi oscillation --- plays a prominent part in quantum optics. From a
fundamental point of view, the existence of discrete Rabi frequencies
provides a direct evidence of electromagnetic field quantization \cite{JC}.
In particular, when there is a dispersion in the photon number, the beating
of different Rabi frequencies gives rise to collapses and revivals in the
inversion probability of the atom \cite{Eberly}. For example, if an atom
initially prepared in its excited state $|e\rangle $ evolves under the
influence of a cavity field, characterized by a photon number distribution
$p_{n}$ and a Rabi frequency $\Omega $, the probability of finding the atom
in the ground state $|g\rangle $ at a later time $t$ is given by
\begin{equation}
P_{eg}(t)=\frac{1}{2}\sum_{n=0}^{\infty }p_{n}\left[ 1-\cos \left( 2\Omega t
\sqrt{n+1}\right) \right] \,,  \label{Rabi}
\end{equation}
which clearly demonstrates the mentioned phenomenon.

\begin{figure}
\includegraphics[height=300pt]{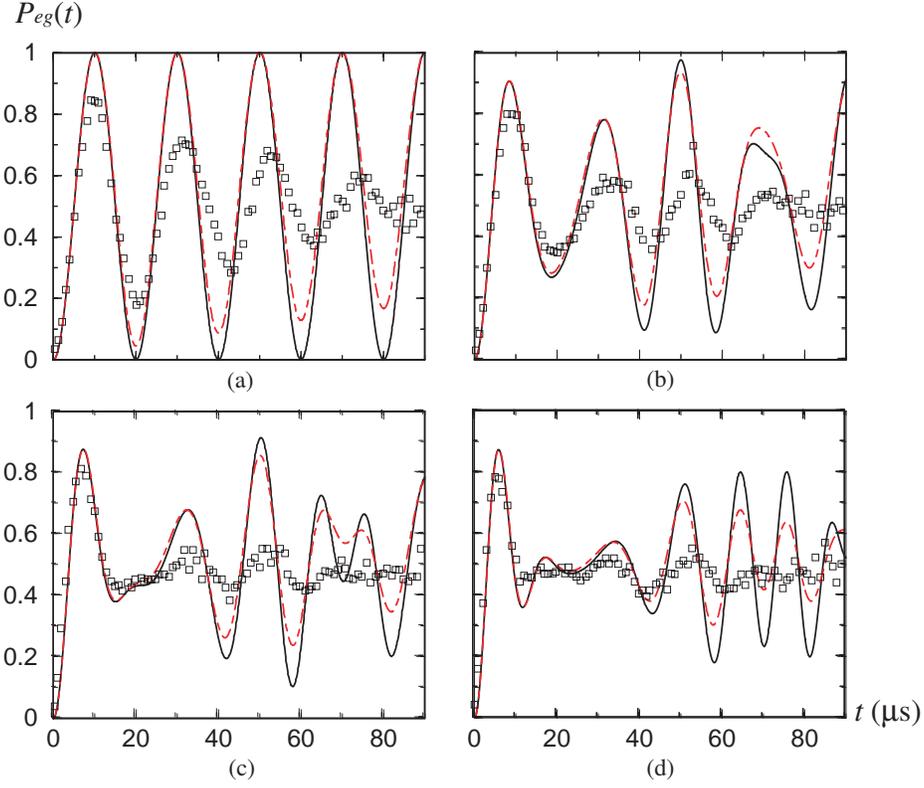}
\caption{$P_{eg}(t)$ vs time for (a) $\overline{n}=0$, (b) $\overline{n}=0.4$,
(c) $\overline{n}=0.85$ and (d) $\overline{n}=1.77$. Boxes represent
experimental data reproduced from Ref.~\cite{MBrune}. Solid lines
are the theoretical results for $\kappa=\gamma=0$ and
$\Omega=\Omega_0=50\pi \, \mathrm{kHz}$. Dot-dashed lines are the
results when finite cavity damping with $\kappa=4.55 \, \mathrm{kHz}$
is taken into account.}
\label{fig1}
\end{figure}

\begin{figure}
\includegraphics[height=300pt]{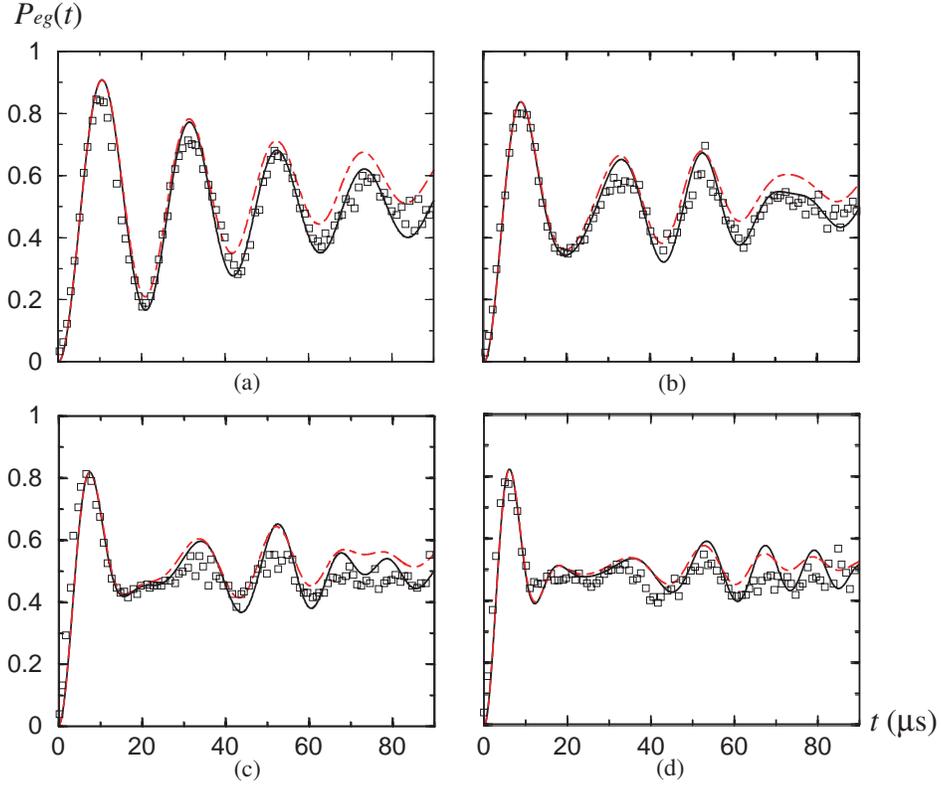}
\caption{$P_{eg}(t)$ vs time for (a) $\overline{n}=0$, (b) $\overline{n}=0.4$,
(c) $\overline{n}=0.85$ and (d) $\overline{n}=1.77$. Boxes represent
experimental data reproduced from Ref.~\cite{MBrune}. Solid lines
are the exact analytic results with collision dephasing $\gamma=19.3 \, \mathrm{kHz}$,
$\kappa =0$ and $\Omega =150.2\,\mathrm{kHz}$.
Dot-dashed lines are the exact numerical results with the same set of
parameters except that $\protect\kappa =4.55\,\mathrm{kHz}$.}
\label{fig2}
\end{figure}

\begin{figure}
\includegraphics[height=300pt]{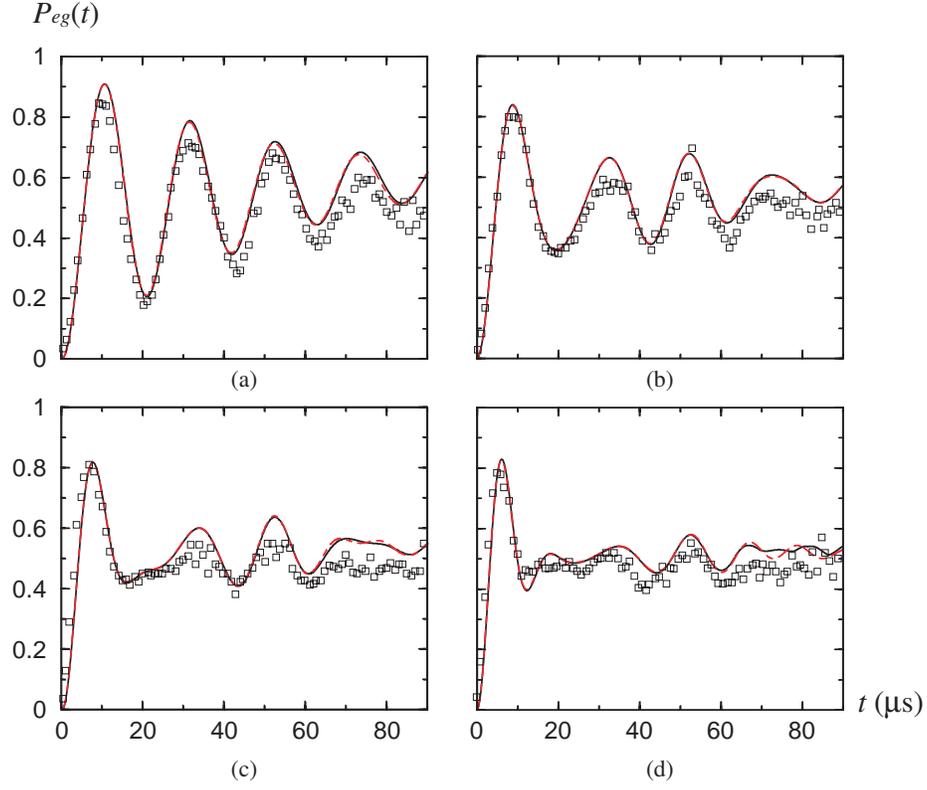}
\caption{$P_{eg}(t)$ vs time for (a) $\overline{n}=0$, (b) $\overline{n}=0.4$,
(c) $\overline{n}=0.85$ and (d) $\overline{n}=1.77$. Boxes represent
experimental data reproduced from Ref.~\cite{MBrune}. Solid lines
are the results of the first order perturbation for a case with collision
dephasing $\gamma=19.3 \, \mathrm{kHz}$, $\kappa=4.55 \,
\mathrm{kHz}$ and $\Omega=150.2 \, \mathrm{kHz}$. Dot-dashed lines are the
exact numerical results.}
\label{fig3}
\end{figure}

\begin{figure}
\includegraphics[height=300pt]{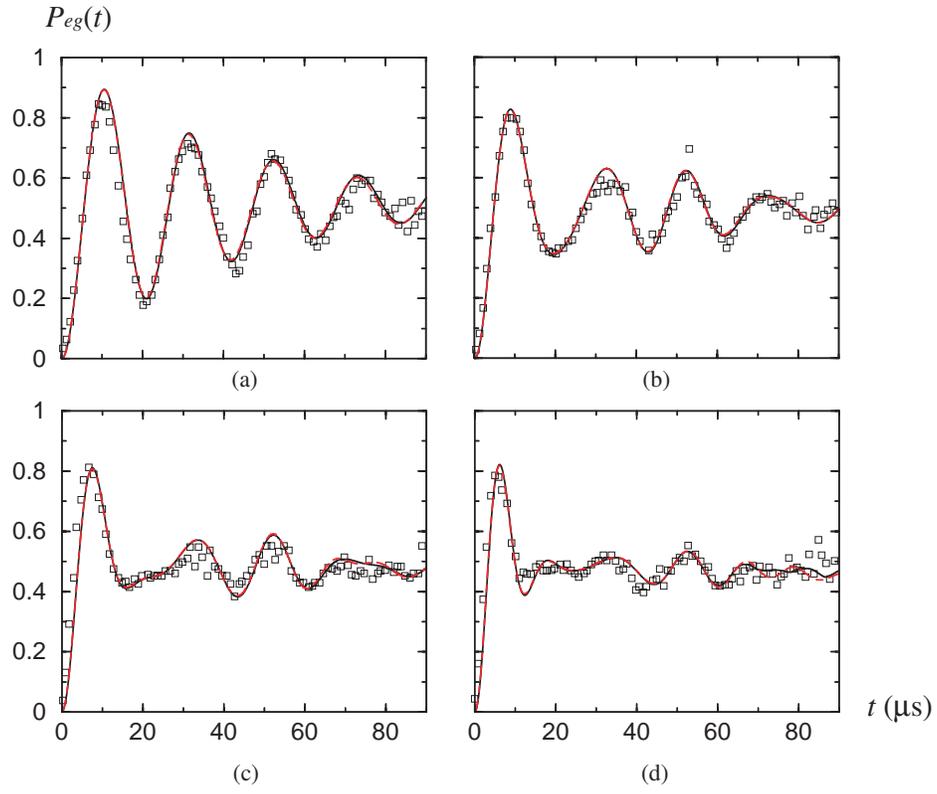}
\caption{Same as Fig.~\ref{fig3}, except the introduction of dark counts
with $\alpha=1.59 \, \mathrm{kHz}$.}
\label{fig4}
\end{figure}

The first observation of quantum Rabi oscillation was made by Rempe \textit{et al}.
some years ago \cite{Rempe}. However, their experiment failed to
obtain a conclusive result due to the limitation on the observation time.
More recently, M.~Brune \textit{et al}. successfully carried out an
experiment to observe the Rabi oscillation \cite{MBrune}, which for the
first time provided direct and clear evidence of field quantization inside a
high-\textit{Q} cavity. In their experiment, Rydberg atoms independently
interact with a photon mode in a superconducting microwave cavity and
undergo transitions between two atomic states with principal quantum numbers
$51$ and $50$ respectively. The \textit{Q}-factor of the cavity mode is $7\times 10^{7}$,
which corresponds to a photon lifetime of $220\,\mu \mathrm{s}$.
The Rabi frequency at the center of the cavity is $\Omega _{0}=50\pi \, \mathrm{kHz}$,
which is sufficiently fast to make Rabi oscillations
observable within the cavity leakage time. The atoms are initially prepared
in the excited state and, in addition to a background $0.8\,\mathrm{K}$
thermal field with a mean photon number $\overline{n}_{\mathrm{th}}\simeq
0.06$, the cavity field is maintained in coherent states with mean photon
number $\overline{n}$ varying from zero to a few photons. The experimental
data of $P_{eg}$ obtained from Ref.~\cite{MBrune}, for $\overline{n}=0$, $0.4
$, $0.85$ and $1.77$ are reproduced here as the boxes in Figs.~\ref{fig1}-\ref{fig4}.
Despite that the data clearly revealed the Rabi oscillations in $P_{eg}(t)$,
the evolution does not conform to that predicted by Eq.~(\ref{Rabi})
(solid lines in Fig.~\ref{fig1}). Instead, it was shown that a best
fit to the data is given by \cite{MBrune}:
\begin{equation}
P_{eg}(t)=\frac{1}{2}\sum_{n=0}^{\infty }p_{n}\left[ 1-\exp (-\Gamma t)\cos
\left( 2\Omega t\sqrt{n+1}\right) \right] \,.  \label{fit}
\end{equation}
The oscillations were damped in such a way that $P_{eg}(t)$ approached $0.5$
in the long term. This behavior is not attributable to cavity leakage, which
would instead lead to $P_{eg}(t)=1$ asymptotically. Furthermore, the photon
lifetime was actually longer than the observation time and the effect of
cavity damping due to leakage should only play a marginal role (see the
dot-dashed lines in Fig.~\ref{fig1}). It was conjectured that decoherence
effect due to collisions with background gas might have contributed to these
damped oscillations \cite{MBrune}.

The aim of this report is to theoretically account for the experimental
results by studying the effects of collision dephasing and cavity leakage.
We will show in the following discussion that these two independent
mechanisms are the main culprits leading to the discrepancy between the
experimental data and the theoretical results given by Eq.~(\ref{Rabi}).

In the rotating wave approximation, the hamiltonian of a two-level atom
interacting with an ideal cavity mode is given by (in units of $\hbar =1$):
\begin{equation}
H_{0}=\frac{\omega _{a}}{2}S_{z}+\omega _{c}a^{\dagger }a+\Omega \left(
a^{\dagger }S_{-}+aS_{+}\right) \,,  \label{H0}
\end{equation}
where $a^{\dagger }$ and $a$ are respectively the creation and annihilation
operators of the cavity field, $S_{z}$ and $S_{\pm }$ are the pseudo-spin
operators of the atomic levels, and $\omega _{a}$ and $\omega _{c}$ are the
atomic transition and cavity mode frequencies respectively. Hereafter, we
will assume exact resonance condition such that $\omega _{a}=\omega _{c}$.
It is worthwhile to note that the vacuum Rabi frequency $\Omega $ in the
hamiltonian can be smaller than the maximum Rabi frequency $\Omega _{0}$
because maximum coupling is attainable only if the two-level atoms are
exactly located at the cavity center \cite{MBrune}. Instead, $\Omega $ will
be considered as a free parameter to fit the leading few Rabi oscillations.

To include the effects of collision dephasing and cavity leakage on the
system, we consider the master equation for the density matrix $\rho $ \cite{Puri,Scully}:
\begin{equation}
\frac{\partial \rho }{\partial t}=-i\left[ H_{0},\rho \right] +L_{c}\rho
+L_{f}\rho \,,  \label{Master}
\end{equation}
where
\begin{equation}
L_{f}\rho =\frac{\kappa }{2}(2a\rho a^{\dagger }-a^{\dagger }a\rho -\rho
a^{\dagger }a)\,,  \label{L1}
\end{equation}
with $\kappa $ being the cavity leakage rate ($\kappa =4.55\,\mathrm{kHz}$
in the experiment), and
\begin{equation}
L_{c}\rho =2\gamma (2S_{z}\rho S_{z}-S_{z}^{2}\rho -\rho S_{z}^{2})\,.
\label{L2}
\end{equation}
The operator $L_{f}$ results from finite leakage at the boundaries of the
cavity, whereas $L_{c}$ describes the effect of collision dephasing
characterized with an energy shift $2\gamma $ \cite{Puri,Scully}. In
general, there is an additional term due to spontaneous decay:
\begin{equation}
L_{s}\rho =-\frac{\Gamma _{e}-\Gamma _{g}}{2}\{S_{z},\rho \}-\frac{\Gamma
_{e}+\Gamma _{g}}{2}\rho \,,  \label{spont}
\end{equation}
with $\Gamma _{e}$ and $\Gamma _{g}$ being the spontaneous decay rates of
the upper and lower energy levels respectively. However, in the experiment
the lifetimes of the two levels are about $30~\mathrm{ms}$ \cite{MBrune},
which are relatively long compared with the atom-cavity interaction time,
and this term is ignored hereafter. Besides, we also assume that the thermal
field outside the cavity is negligible. In terms of the density matrix,
$P_{eg}(t)$ is given by
\begin{equation}
P_{eg}(t)=\sum\limits_{n}\langle n,g|\rho |n,g\rangle \,.  \label{Peg-rho}
\end{equation}

In the current situation, the leakage of the cavity is small. It is
legitimate to take $L_{f}\rho $ as the small term and Eq.~(\ref{Master}) can
then be solved by method of perturbation. First, the density matrix is
written as
\begin{equation}
\rho =\rho ^{(0)}+\rho ^{(1)}\,,  \label{pert}
\end{equation}
where $\rho ^{(1)}$ is a small correction to the zeroth order density matrix
$\rho ^{(0)}$. It is then readily shown that
\begin{equation}
\frac{\partial \rho ^{(0)}}{\partial t}=-i\left[ H_{0},\rho ^{(0)}\right]
+L_{c}\rho ^{(0)}\,,  \label{pert0}
\end{equation}
and
\begin{equation}
\frac{\partial \rho ^{(1)}}{\partial t}=-i\left[ H_{0},\rho ^{(1)}\right]
+L_{c}\rho ^{(1)}+L_{f}\rho ^{(0)}\,.  \label{pert1}
\end{equation}
Secondly, we solve Eq.~(\ref{pert0}) for the initial conditions:
\begin{eqnarray*}
\left\langle n,e|\rho ^{(0)}|n,e\right\rangle  &=&p_{n}\,, \\
\left\langle n,g|\rho ^{(0)}|n,g\right\rangle  &=&\left\langle n,e|\rho
^{(0)}|n+1,g\right\rangle =\left\langle n+1,g|\rho ^{(0)}|n,e\right\rangle
=0\,.
\end{eqnarray*}
When the field is prepared in a coherent state, the photon distribution is
Poissonian with $p_{n}$ given by:
\begin{equation}
p_{n}=e^{-\overline{n}}\frac{\overline{n}^{n}}{n!}\,.  \label{Po}
\end{equation}
The explicit solution to Eq.~(\ref{pert0}) is:
\begin{equation}
\left\langle 0,g|\rho ^{(0)}|0,g\right\rangle =0\,,  \label{sol0-1}
\end{equation}
\begin{equation}
\left\langle n,g|\rho ^{(0)}|n,g\right\rangle =\frac{1}{2}p_{n-1}\left[
1-\exp (-\gamma t)\frac{\cos (\lambda _{n-1}t-\phi _{n-1})}{\cos \phi _{n-1}}
\right] \ ,  \label{sol0-2}
\end{equation}
\begin{equation}
\left\langle n,e|\rho ^{(0)}|n,e\right\rangle =\frac{1}{2}p_{n}\left[ 1+\exp
(-\gamma t)\frac{\cos (\lambda _{n}t-\phi _{n})}{\cos \phi _{n}}\right] \,,
\label{sol0-3}
\end{equation}
\begin{equation}
\left\langle n,e|\rho ^{(0)}|n+1,g\right\rangle =\frac{i}{2}p_{n}\exp
(-\gamma t)\ \frac{\sin \lambda _{n}t}{\cos \phi _{n}}\,,  \label{sol0-4}
\end{equation}
where
\begin{equation}
\lambda _{n}=\sqrt{4(n+1)\Omega ^{2}-\gamma ^{2}}  \label{lambda_n}
\end{equation}
and
\begin{equation}
\tan \phi _{n}=\frac{\gamma }{\lambda _{n}}\,.  \label{phi_n}
\end{equation}
Eq. (\ref{Peg-rho}) can hence be expressed explicitly as
\begin{equation}
P_{eg}(t)=\frac{1}{2}\sum\limits_{n=0}^{\infty }p_{n}\left[ 1-\exp (-\gamma
t)\frac{\cos (\lambda _{n}t-\phi _{n})}{\cos \phi _{n}}\right] \,.
\label{Peg-sol0}
\end{equation}

From the results, we see that the mechanism of collision dephasing gives
rise to remarkable effects. The Rabi oscillations are damped with modified
frequencies given by Eq.~(\ref{lambda_n}); and there is also a phase shift
as given by Eq.~(\ref{phi_n}). The solid-lines in Fig.~\ref{fig2} show the
zeroth order perturbation results for $\Omega _{0}=150.2\,\mathrm{kHz}$ and
$\gamma =19.3\ \mathrm{kHz}$, which is observed to provide best fits to the
four sets of experimental data. The results obtained by numerically solving
the exact master equation (i.e. Eq.~(\ref{Master})) are also shown there as
dot-dashed lines. Interestingly enough, there is a fairly good apparent
agreement between the zeroth order results and the experimental data. We
consider this as a coincidence due to two counter-balancing effects, namely
the effects of cavity leakage and dark counts in the detector used in the
experiment \cite{MBrune}, which have been neglected in the forgoing
discussion. We will return to this point later.

To obtain a better agreement between the perturbative and the numerical
results, the first order calculation is carried out, yielding the results:
\begin{equation}
\left\langle 0,g|\rho ^{(1)}|0,g\right\rangle =-\frac{\kappa }{4\Omega }p_{0}
\left[ 2\sin \phi _{0}-2\Omega t+\exp (-\gamma t)\frac{\sin (\lambda
_{0}t-2\phi _{0})}{\cos \phi _{0}}\right] \,,  \label{sol1-1}
\end{equation}
\begin{eqnarray}
&&\left\langle n,g|\rho ^{(1)}|n,g\right\rangle   \nonumber \\
&=&\kappa \frac{\sin 2\phi _{n}}{4n\lambda _{n}}p_{n-1}+\frac{1}{4}\left[
(2n+1)p_{n}-(2n-1)p_{n-1}\right] \kappa t  \nonumber \\
&&-\kappa \exp (-\gamma t){\Bigg\{}(2n^{2}+n+1)p_{n}\frac{\sin 2\phi _{n-1}}{%
4\lambda _{n-1}}\cos \lambda _{n-1}t  \nonumber \\
&&-\frac{(2n-1)(1-\sqrt{n})p_{n-1}+[4n^{2}+3n+1-2(4n^{2}+2n+1)\sin ^{2}\phi
_{n-1}]p_{n}}{4\lambda _{n-1}}\sin \lambda _{n-1}t  \nonumber \\
&&+\frac{n(n+1)p_{n}}{2\lambda _{n}}\sin \lambda _{n}t+\frac{%
(2n^{2}+3n+2)p_{n}}{4\lambda _{n}}\sin (\lambda _{n}t-2\phi _{n})  \nonumber
\\
&&-\frac{(2n-1)\sqrt{n}p_{n-1}}{2\lambda _{n-1}}\Omega t\cos (\lambda
_{n-1}t-\phi _{n-1}){\Bigg\}}\,,  \label{sol1-2}
\end{eqnarray}
\begin{eqnarray}
&&\left\langle n,e|\rho ^{(1)}|n,e\right\rangle   \nonumber \\
&=&\kappa \left( \frac{\sin 2\phi _{n}}{2\lambda _{n}}p_{n}-\frac{(2n+3)\sin
2\phi _{n+1}}{4(n+1)\lambda _{n+1}}p_{n+1}\right) +\frac{1}{4}\left[
(2n+3)p_{n+1}-(2n+1)p_{n}\right] \kappa t  \nonumber \\
&&+\kappa \exp (-\gamma t){\Bigg\{}\frac{(2n^{2}+5n+4)p_{n+1}-2p_{n}}{%
4\lambda _{n}}\sin 2\phi _{n}\cos \lambda _{n}t  \nonumber \\
&&+\frac{\left[ (2n+1)(\sqrt{n+1}-1)+2\cos 2\phi _{n}\right] p_{n}-\left[
n+1+(4n^{2}+10n+7)\cos 2\phi _{n}\right] p_{n+1}}{4\lambda _{n}}\sin \lambda
_{n}t  \nonumber \\
&&+\frac{(n+1)(n+2)p_{n+1}}{2\lambda _{n+1}}\sin \lambda _{n+1}t+\frac{%
(2n^{2}+7n+5)p_{n+1}}{4\lambda _{n+1}}\sin (\lambda _{n+1}t-2\phi _{n+1})
\nonumber \\
&&-\frac{(2n+1)\sqrt{n+1}p_{n}}{2\lambda _{n}}\Omega t\cos (\lambda
_{n}t-\phi _{n}){\Bigg\}}\,,  \label{sol1-3}
\end{eqnarray}
\begin{eqnarray}
&&\left\langle n,e|\rho ^{(1)}|n+1,g\right\rangle   \nonumber \\
&=&\frac{\kappa }{8\sqrt{n+1}\Omega }\left( p_{n}-p_{n+1}\right)
-\frac{\kappa }{8\sqrt{n+1}\Omega }\exp (-\gamma t){\Bigg\{}\left[
p_{n}-(3n^{2}+8n+6)p_{n+1}\right] \cos \lambda _{n}t  \nonumber \\
&&+\left[ p_{n}+(n^{2}+4n+2)p_{n+1}\right] \tan \phi _{n}\sin \lambda
_{n}t+2(n+1)(3n+5)p_{n+1}\cos \lambda _{n+1}t  \nonumber \\
&&+2(n+1)^{2}p_{n+1}\tan \phi _{n+1}\sin \lambda _{n+1}t+4(n+1)(2n+1)p_{n}
\frac{\Omega ^{2}}{\lambda _{n}}t\sin \lambda _{n}t{\Bigg\}}\,.
\label{sol1-4}
\end{eqnarray}

The first order results with the same parameters are shown in Fig.~\ref{fig3}.
In the time regime under consideration, the first order results give
excellent approximation to the exact numerical solution that includes the
leakage effect. However, as shown in Fig.~\ref{fig3}, the introduction of
leakage obviously also worsens the agreement between our theoretical results
and the experimental data. As suggested in Ref.~\cite{MBrune}, we propose that
the effect of cavity damping might have been counter-balanced by dark counts
in the atomic detector, which become increasingly important at long times
because of low atomic fluxes. If it is assumed that within the time range of
the experiment, the detection rate approximately goes exponentially with $t$,
an extra factor $e^{-\alpha t}$ should be introduced. In other words,
\begin{equation}
\lbrack P_{eg}(t)]_{\mathrm{experiment}}=e^{-\alpha t}[P_{eg}(t)]_{\mathrm{theory}}\;.
\end{equation}
In the time regime under consideration, this is roughly the same as assuming
a linear time dependence of the dark count rate. In order to counter-balance
the effect of cavity damping, we find that $\alpha $ should assume a value
around $1.59\,\mathrm{kHz}$. In other words, the proportion of dark counts
should increase from $0$ at $t=0$ to around $13\%$ at $t=90\,\mu \mathrm{s}$.
In Fig.~\ref{fig4}, the overlying solid lines and dot-dashed lines
respectively show the first-order perturbation and numerical results with
the effect of dark counts taken into account. It clearly shows an excellent
agreement with experimental data.

\begin{figure}
\includegraphics[height=300pt]{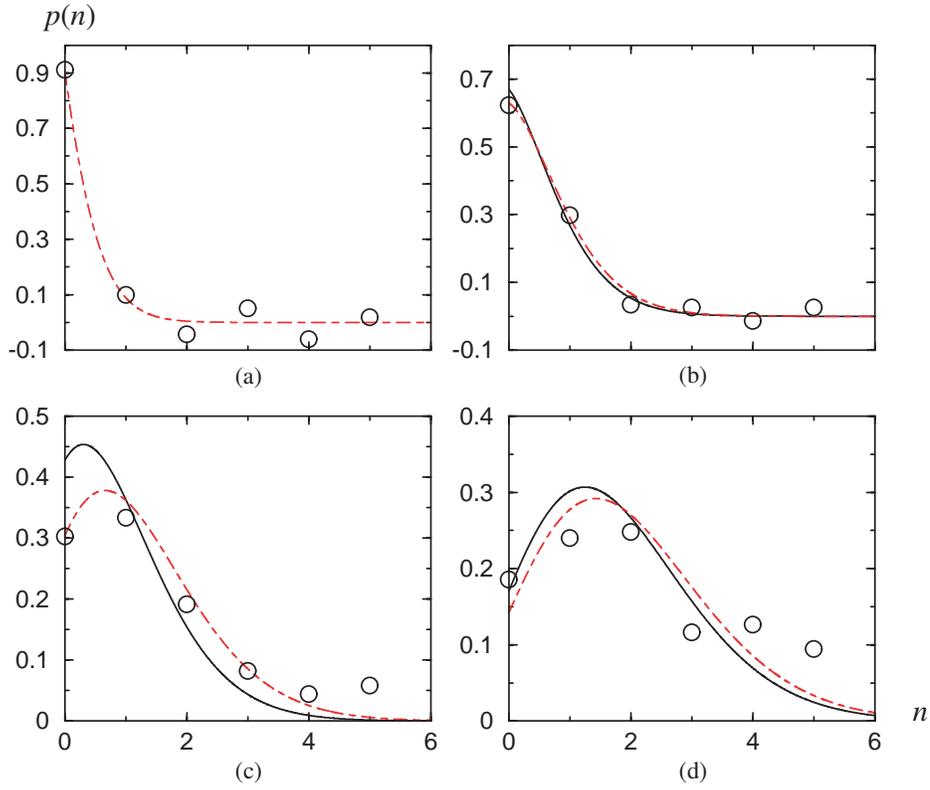}
\caption{Photon statistics for the four experimental situations. Circles
represent the results obtained by inverting the optimal $\tilde{p}_n$ from
the experimental data. Dot-dashed lines are Poissonian distributions that
best fit the circles. Solid lines are the Poissonian distributions specified
in Ref.~\cite{MBrune}. }
\label{fig5}
\end{figure}

Moreover, the results of the first order perturbation allow us to write
Eq.~(\ref{Peg-rho}) in the form
\begin{equation}
P_{eg}(t)=\sum_{n=0}^{\infty }p_{n}f_{n}(t)\,,  \label{fn}
\end{equation}
where the functions $f_{n}(t)$ are obtained by grouping terms in
Eqs.~(\ref{sol0-1})-(\ref{sol0-4}) and (\ref{sol1-1})-(\ref{sol1-4}). Instead of
evaluating $P_{eg}(t)$ from the given photon distribution function $p_{n}$ \cite{MBrune},
we can in fact search for initial states that best fit the
data. With the set of functions $f_{n}(t)$, we perform a least-square best
fit to obtain the optimal photon distribution $\tilde{p}(n)$. The results of
this inversion process are shown by circles in Fig.~\ref{fig5}, in which the
solid lines represent the theoretical results for the initial states given
in Ref.~\cite{MBrune}, namely $\overline{n}=0$, $0.4\pm 0.02$, $0.85\pm 0.04$ and
$1.77\pm 0.15$. It is observed that the results we obtained show a fairly
good agreement with the theoretical values. Furthermore, we perform
independent best-fits on $p(n)$ to find the coherent states that give the
best agreement. We found that the distributions are respectively best fit by
coherent states with $\overline{n}=0.098$, $0.46$, $1.19$ and $1.95$,
signifying the agreement of the experimental and theoretical results.

In conclusion, we have theoretically analyzed the evolution of a
single atom inside a high-\textit{Q} cavity in the presence of
collision dephasing and leakage mechanisms, and the results are
compared with experimental observations performed by Brune
\textit{et al}. \cite{MBrune}. Collision dephasing, leakage and
dark counts are shown to be the major factors affecting the
experimental results. Interestingly enough,  cavity leakage and
dark counts tend to produce opposite effects on the inversion
probability. By assuming suitable values of collision Stark shift
and dark count rate, we have explicitly demonstrated that the
experimental data agree nicely with the theoretical prediction.

\begin{acknowledgments}
We thank G.~Q.~Ge and C.~K.~Law respectively for drawing our attention to
Brune \textit{et al.}'s experiment and discussions. Our work is supported in
part by the Hong Kong Research Grants Council (grant No: CUHK4282/00P) and a
direct grant (Project ID: 2060150) from the Chinese University of Hong Kong.
\end{acknowledgments}


\newpage

\newpage

\newpage

\end{document}